\documentclass[10pt,twocolumn,letterpaper]{article}

\usepackage{iccv}
\usepackage{times}
\usepackage{epsfig}
\usepackage{graphicx}
\usepackage{amsmath}
\usepackage{amssymb}
\usepackage{multirow}%

\usepackage[breaklinks=true,bookmarks=false]{hyperref}

\iccvfinalcopy 

\def\iccvPaperID{****} 

\ificcvfinal\fi

\begin{document}

\title{Enhancing Medical Image Segmentation: Optimizing Cross-Entropy Weights and Post-Processing with Autoencoders}

\author{Pranav Singh\\
Tandon School of Engineering\\
New York University\\
{\tt\small ps4364@nyu.edu}
\and
Luoyao Chen\\
Center for Data Science \\
New York University\\
{\tt\small lc4866@nyu.edu}
\and
Mei Chen\\
Center for Data Science \\
New York University\\
{\tt\small mc8895@nyu.edu}
\and
Jinqian Pan\\
Center for Data Science \\
New York University\\
{\tt\small jp6218@nyu.edu}
\and
Raviteja Chukkapalli\\
Courant Institute of Mathematical Sciences \\
New York University\\
{\tt\small rc5124@nyu.edu}
\and
Shravan Chaudhari \\
Courant Institute of Mathematical Sciences \\
New York University\\
{\tt\small shravan.c@nyu.edu}
\and
Jacopo Cirrone\\
Center for Data Science \\
New York University\\
{\tt\small cirrone@courant.nyu.edu}
}

\maketitle
\ificcvfinal\thispagestyle{empty}\fi

\begin{abstract}
   The task of medical image segmentation presents unique challenges, necessitating both localized and holistic semantic understanding to accurately delineate areas of interest, such as critical tissues or aberrant features. This complexity is heightened 
   in medical image segmentation due to the high degree of inter-class similarities, intra-class variations, and possible image obfuscation. The segmentation task further diversifies when considering the study of histopathology slides for autoimmune diseases like dermatomyositis. The analysis of cell inflammation and interaction in these cases has been less studied due to constraints in data acquisition pipelines. Despite the progressive strides in medical science, we lack a comprehensive collection of autoimmune diseases. As autoimmune diseases globally escalate in prevalence and exhibit associations with COVID-19, their study becomes increasingly essential. While there is existing research that integrates artificial intelligence in the analysis of various autoimmune diseases, the exploration of dermatomyositis remains relatively underrepresented. In this paper, we present a deep-learning approach tailored for 
   Medical image segmentation. Our proposed method outperforms the current state-of-the-art techniques by an average of 12.26\% for U-Net and 12.04\% for U-Net++ across the ResNet family of encoders on the dermatomyositis dataset. Furthermore, we probe the importance of optimizing loss function weights and benchmark our methodology on three challenging medical image segmentation tasks.
\end{abstract}

\section{Introduction} \label{intro}

\begin{figure}[!ht]
\centering
\includegraphics[width=0.1\textwidth,height=0.2\linewidth]{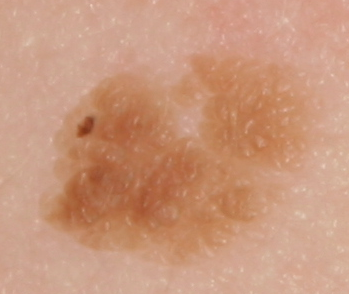}
\includegraphics[width=0.1\textwidth,height=0.2\linewidth]{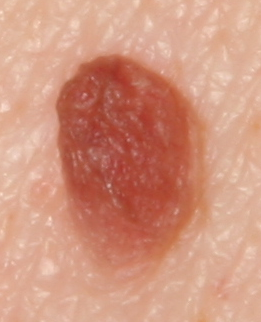}
\includegraphics[width=0.1\textwidth,height=0.2\linewidth]{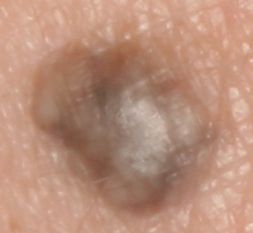}
\includegraphics[width=0.1\textwidth,height=0.2\linewidth]{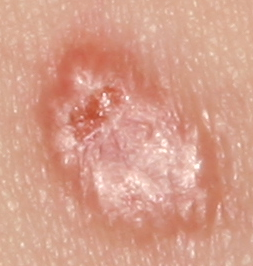}
\newline
\caption{Samples from the dermofit dataset for segmentation, we observe a large lesion color variability from left to right.  Interestingly, the background color remains relatively consistent throughout the samples. We also observe a change in contrast from left to right. These confounding factors make the segmentation of skin lesions difficult.}
\label{fig:var_ex}
\end{figure}

\begin{figure}[!ht]
\centering
\includegraphics[width=0.15\textwidth,height=0.2\linewidth]{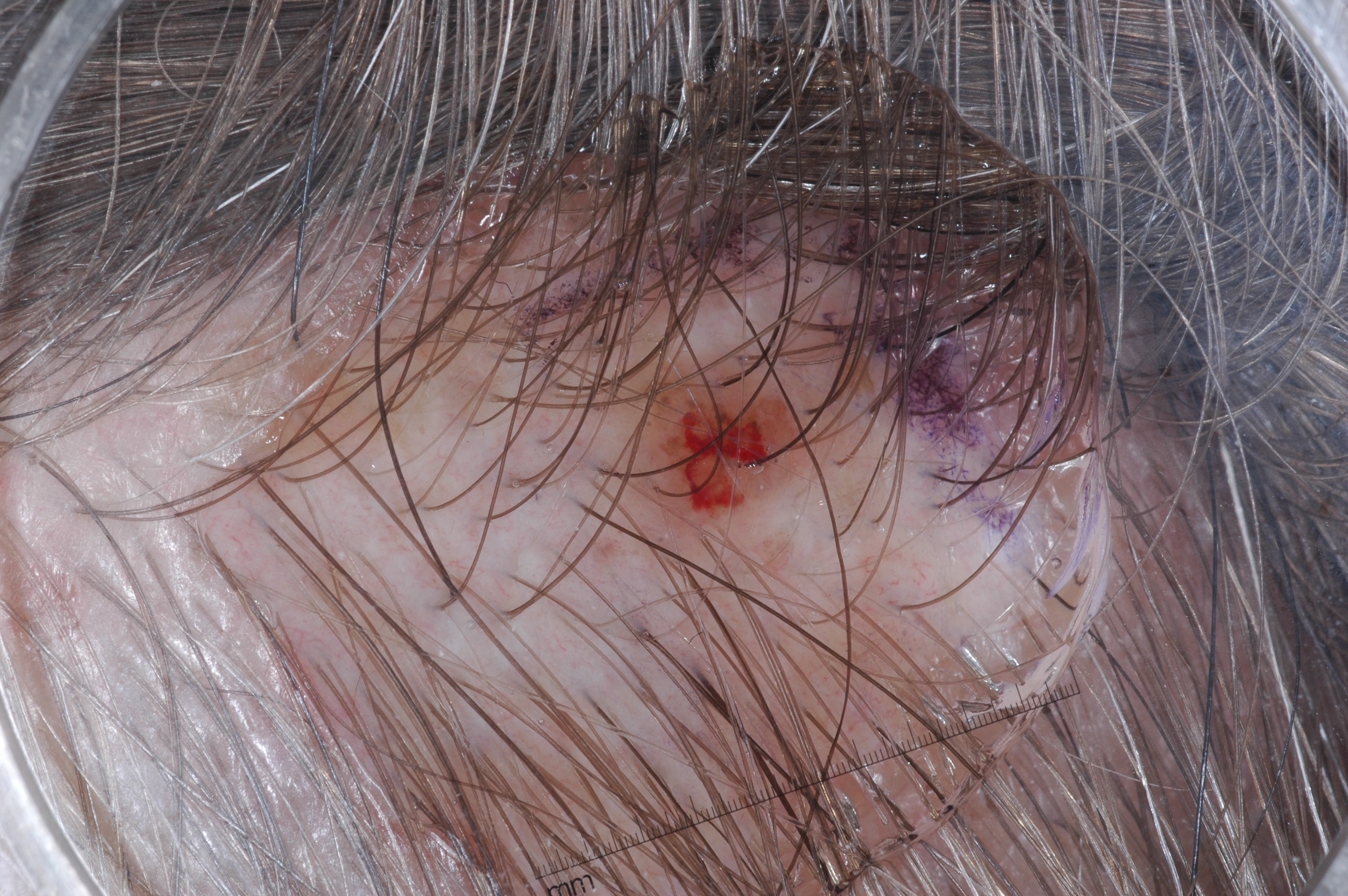}
\includegraphics[width=0.15\textwidth,height=0.2\linewidth]{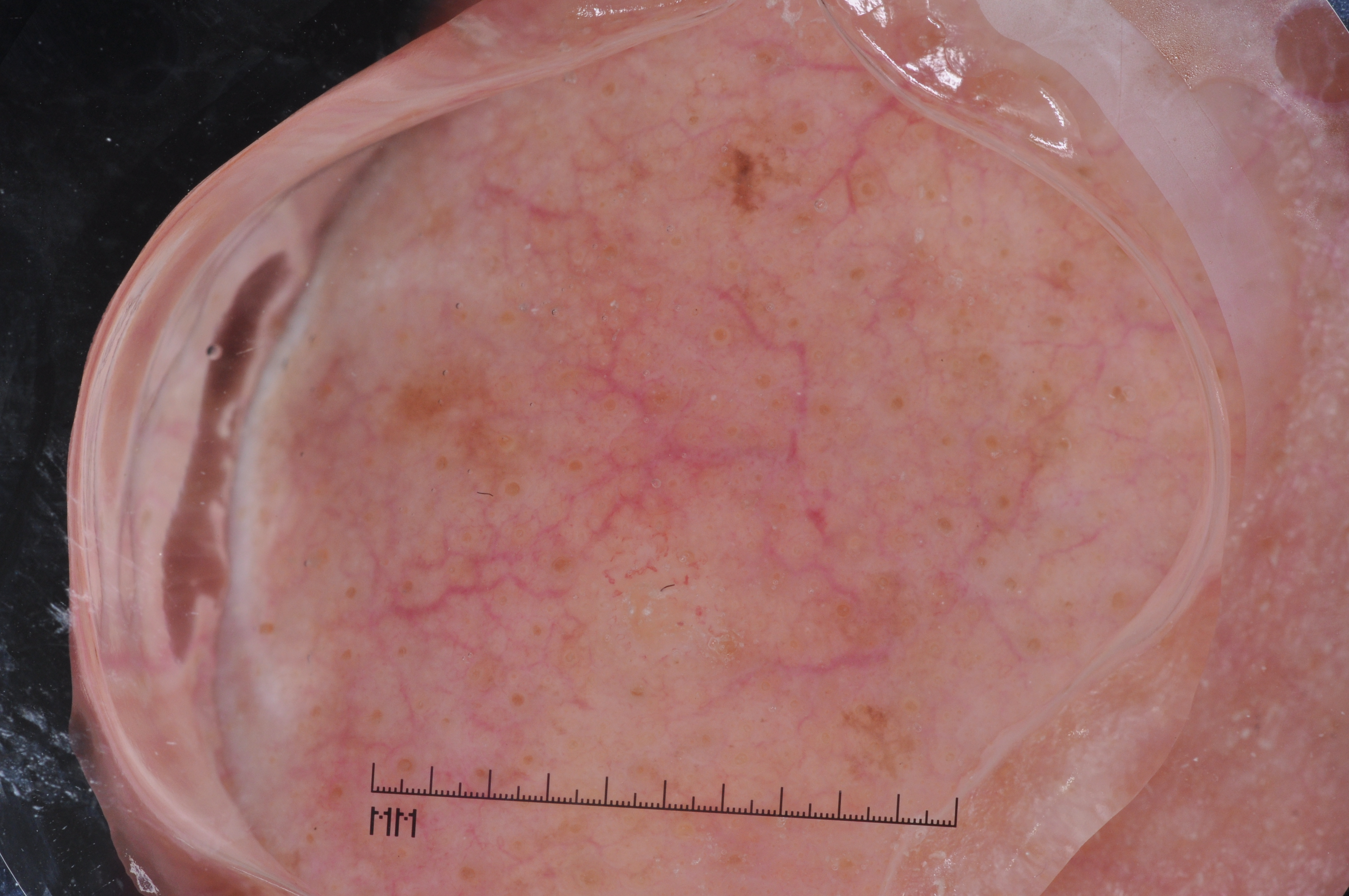}
\includegraphics[width=0.15\textwidth,height=0.2\linewidth]{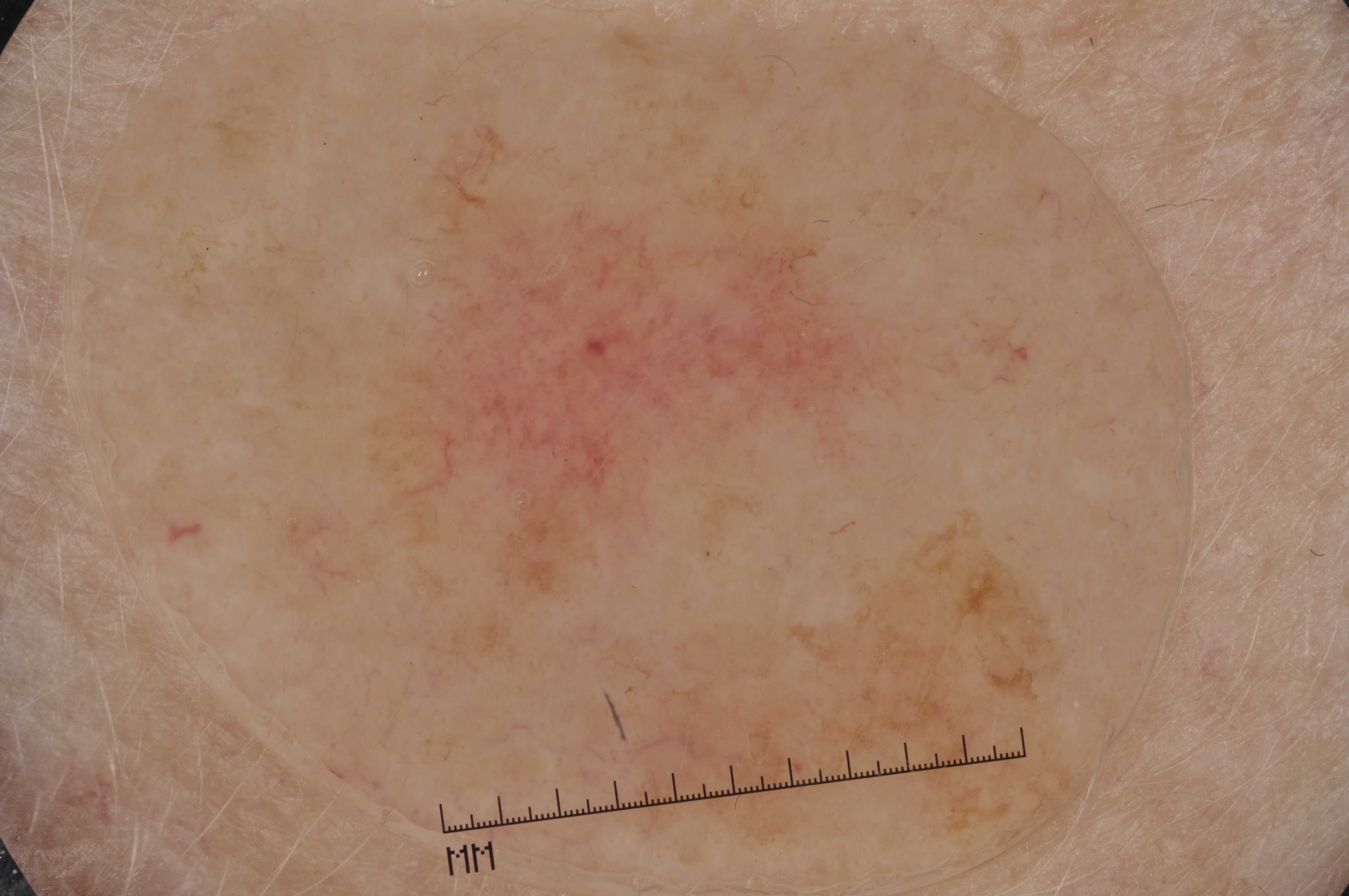}

\caption{This figure contains samples from the ISIC 2017 dataset. The ISIC-2017 dataset exhibits obscuration and significant intra-class variability, resembling the Dermofit dataset (from Figure \ref{fig:var_ex}) regarding inter-class similarity and low contrast (rightmost image). }
\label{fig:obs_ex}
\end{figure}

\begin{figure}[!h]
    \centering
    \includegraphics[width=0.2\textwidth,height=0.2\textwidth]{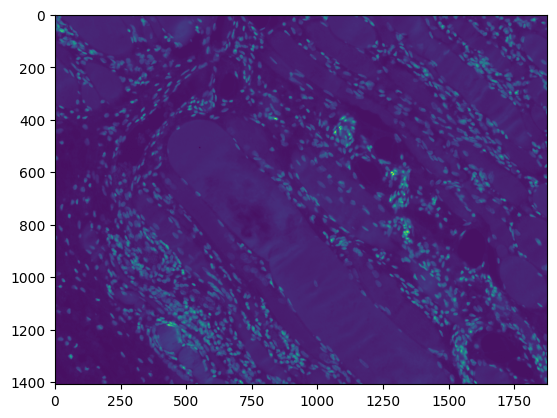}
    \includegraphics[width=0.2\textwidth,height=0.2\textwidth]{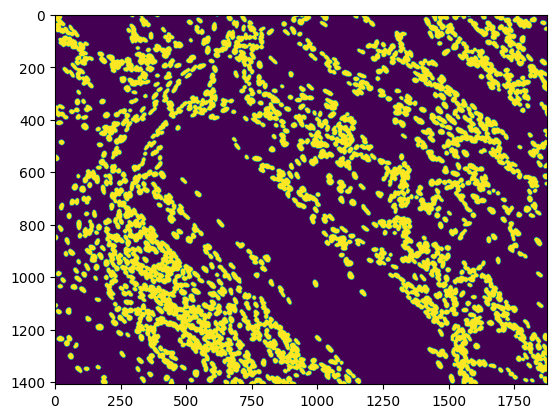}

    \includegraphics[width=0.2\textwidth,height=0.2\textwidth]{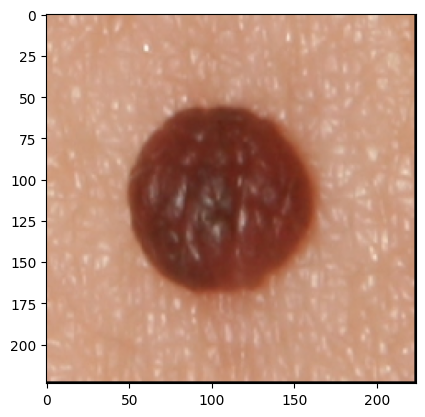}
    \includegraphics[width=0.2\textwidth,height=0.2\textwidth]{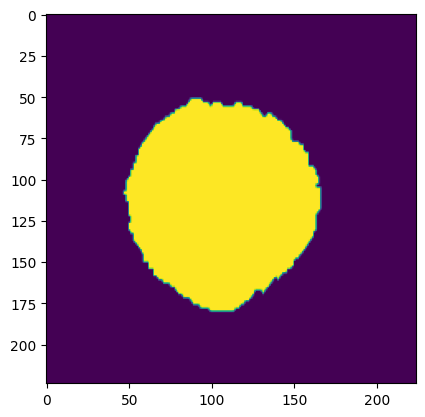}

    \includegraphics[width=0.2\textwidth,height=0.2\textwidth]{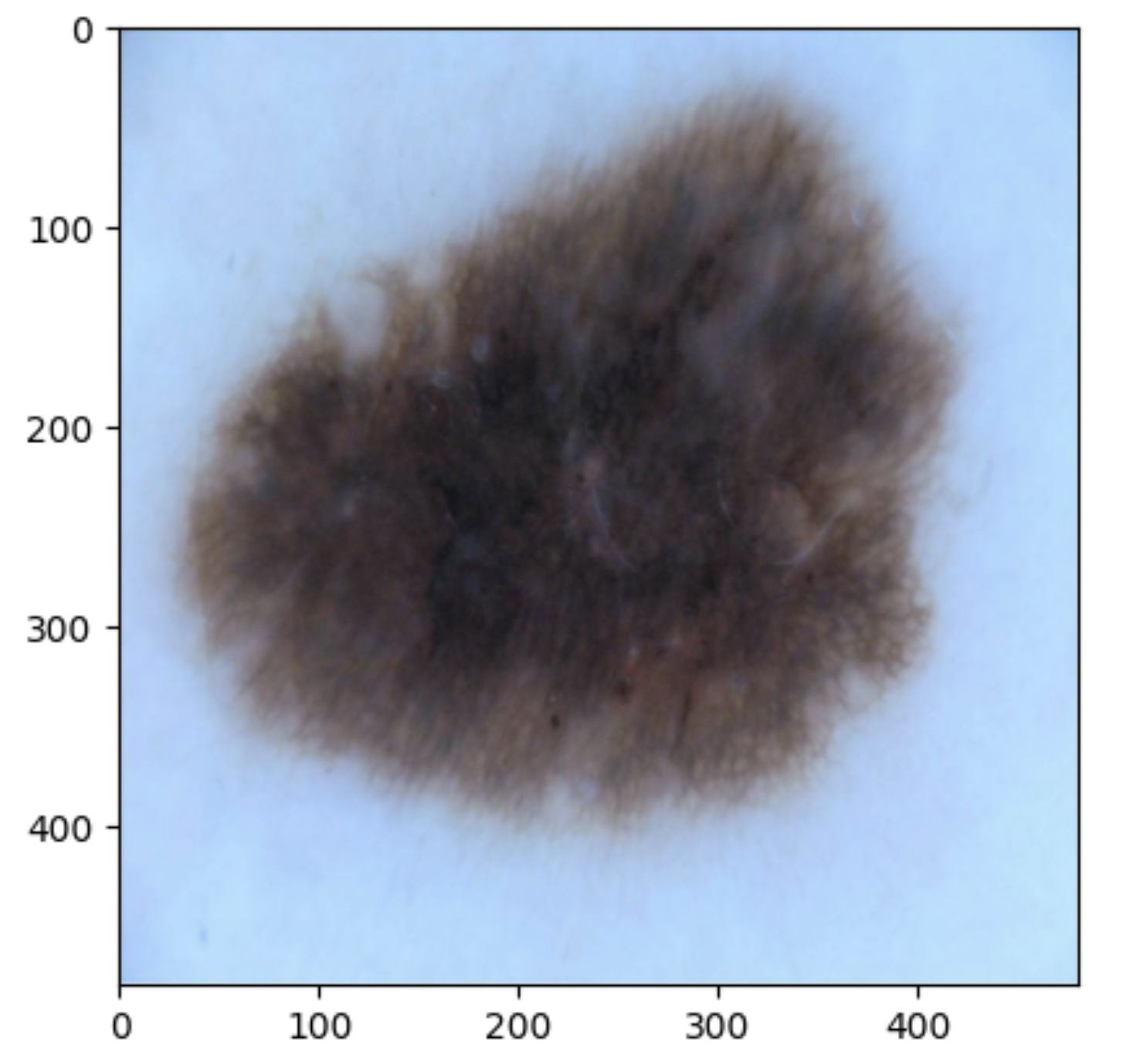}
    \includegraphics[width=0.2\textwidth,height=0.2\textwidth]{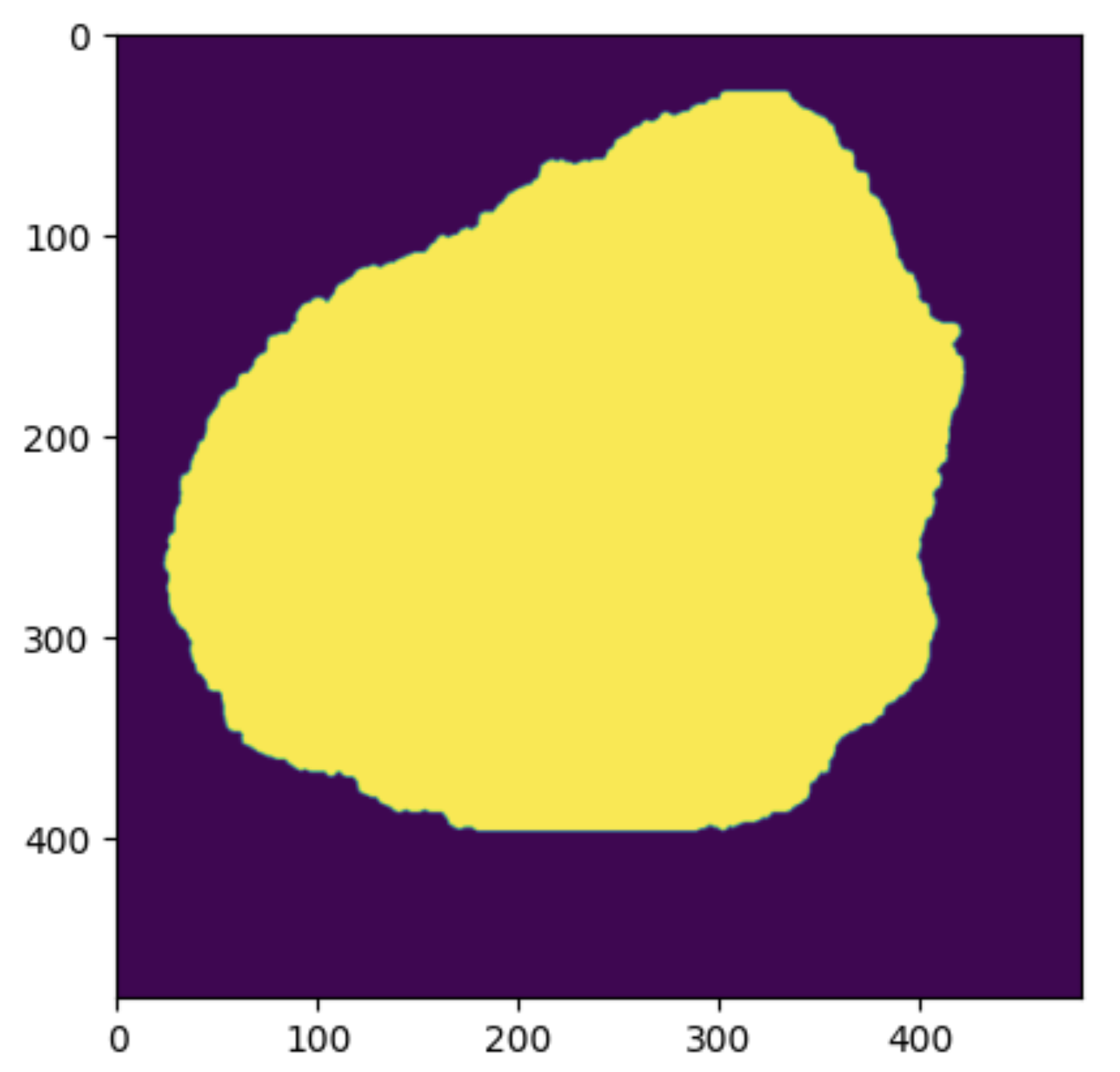}
\caption{Semantic segmentation task as defined in Section \ref{methods} with input image on the left and the corresponding ground truth on the right. On top, we have a sample image (on the left) and the corresponding ground truth (on the right) from the dermatomyositis dataset. Similarly, a sample from the dermofit dataset is in the middle, and a sample from the ISIC-2017 dataset is at the bottom. Unlike a single blob in the skin lesion dataset samples from dermofit and ISIC 2017, we observe that the histopathology whole slide image has many more fine-grained objects with hard-to-segment boundaries. For all the images on the right, the yellow area represents the region of interest (foreground), and the rest is the area other than the region of interest (background).}
\label{fig:isic_data}
\end{figure}

The development of potent CAD (Computer Aided Diagnosis) strategies has been aided by advances in computational power and image analysis algorithms over the past decade. Medical imaging is fundamental to these CAD methods. Obtaining accurate results from CAD techniques relies on acquiring high-quality medical imaging and corresponding annotation. These CAD approaches facilitate various tasks such as image classification, segmentation, spatial mapping, and tracking. Out of these, medical image segmentation is a particularly challenging task due to several complexities. For example, in skin lesion image segmentation, there exists significant intra-class variability and inter-class similarity. This issue is exacerbated by the presence of obscuration and low contrast, which makes the task of separating the affected area from the surrounding image more challenging. On the other hand, sometimes the data required to segment is very complex, with multiple fine-grained and hard-to-segment objects, for example, in the case of histopathology data of dermatomyositis (a kind of autoimmune disease). We provide a few examples of the large variability and low contrast in Figure \ref{fig:var_ex}, obscuration in Figure \ref{fig:obs_ex}, and multiple hard-to-segment small objects in Figure \ref{fig:isic_data}.

\begin{figure*}[!h]
    \centering
    \includegraphics[width=0.8\textwidth]{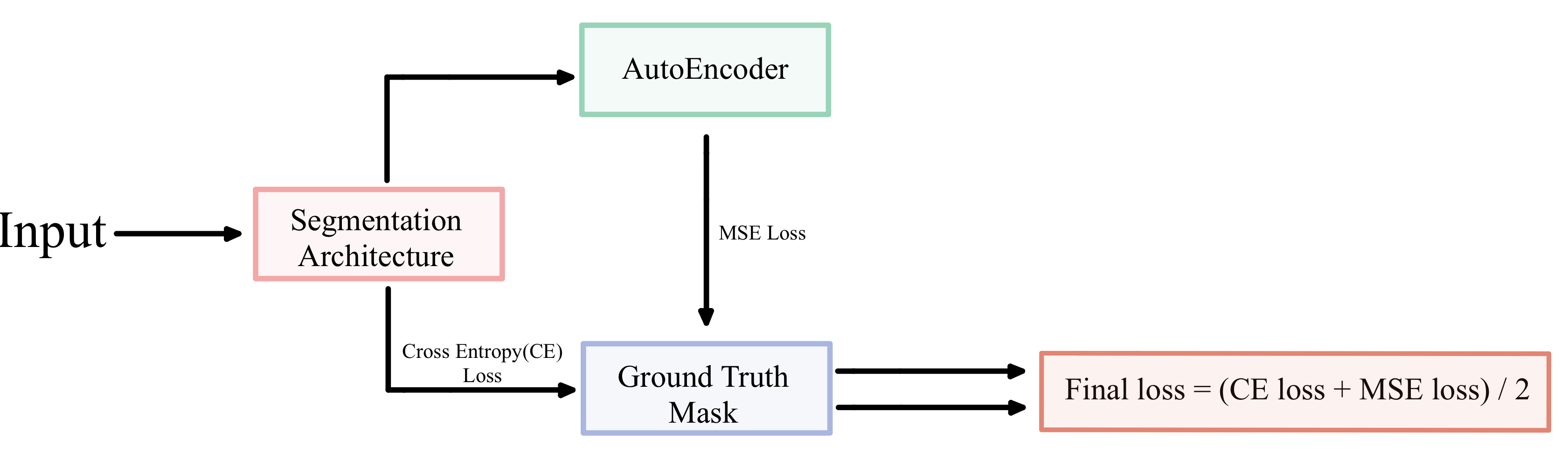}
    \caption{DEDL Architecture\cite{singh2022data} with the APP (Autoencoder Post Processing) for dermatomyositis image segmentation as described in Section \ref{ai-in-ai}. We use this architecture as our baseline and propose changes to it to improve by around 12.26\% for U-Net and 12.04\% for U-Net++ - this is a considerable improvement over DEDL, as DEDL improved over the previous state-of-the-art approach Van Buren \etal \cite{van2022artificial} by around 5\% for segmentation.} 
    \label{fig:DEDL}
\end{figure*}

In addition to these modality-specific complexities, medical imaging datasets are considerably smaller than natural datasets. The main reason for this is the significant expenses and time involved in gathering, annotating medical datasets and privacy concerns. Medical imaging datasets can only be labeled by highly specialized clinicians instead of the possibility of crowdsourced labeling in the case of natural datasets. Privacy concerns pose significant challenges in the open sourcing of medical datasets, particularly for rare or emerging diseases. Medical datasets are typically restricted to institutional use, even when made available \cite{matsoukas2022makes}.

Despite significant progress in medical science, some diseases have not yet been fully comprehended \cite{cooper2009recent}. Autoimmune diseases are a notable category in this context. 
The lack of a comprehensive catalog of autoimmune diseases, unlike other diseases, is attributed to the diverse nature of their onset and progression \cite{stafford2020systematic}. There are still important research questions for autoimmune diseases regarding environmental triggers, pathogenesis, cell inflammation, and interaction.  Currently, there are over 80 classified autoimmune diseases.  Immune-modulatory drugs are commonly employed for the treatment of autoimmune diseases. However, these drugs have a wide range of effects and lack specificity for autoimmune diseases. Unfortunately, their usage is often linked to other infections and malignant diseases as undesirable side effects. Patients often have limited or no response to these treatments due to the variability within these disorders. So, there is a pressing need for more advanced, fast, and accurate ways to find novel relationships and pathologies that can lead to more effective treatments for autoimmune diseases. To accomplish this, it is imperative to develop precise and adaptable techniques for analyzing autoimmune diseases related medical images. Implementing AI-based Computer-Aided Diagnosis (CAD) is a potential strategy for achieving this objective. In contrast to other diseases, however, lacking a definitive list and a limited  understanding of autoimmune disorders presents a challenge. Consequently, there are few established data collection mechanisms for autoimmune diseases. These factors contribute to the paucity of research on the intersection of autoimmune diseases and CAD approaches. Most extant research in this field is either outdated or lacks open-source methodologies. The study of autoimmune diseases is paramount due to their increasing prevalence\cite{ehrenfeld2020covid,galeotti2020autoimmune,lerner2015world}. Autoimmune diseases impact a significant portion of the global population, ranging from 5\% to 8\%. These conditions cause considerable distress to patients and have been found to have connections with COVID-19, the primary cause of the recent worldwide pandemic \cite{sharma2023high,sriram2021covid}. To bridge the divide, Van Buren \etal \cite{van2022artificial} and Singh \& Cirrone \cite{singh2022data} have made attempts. The main focus of these studies is dermatomyositis. This rare autoimmune disease has received limited attention at the intersection of medical imaging and the application of AI (Artificial Intelligence) for medical image analysis. With this paper,

\begin{itemize}
  \item We improve upon the existing state-of-the-art approach \cite{singh2022data} for dermatomyositis segmentation by an average of 12.26\% for U-Net and 12.04\% for Unet++ in Section \ref{result_1}. Additionally, we benchmark our approach on two other challenging skin-related datasets.
  \item We study the impact of adding a post-processing autoencoder in addition to U-Net and U-Net++ on three medical imaging datasets in Section \ref{subsec3}.
  \item We investigate the significance of cross-entropy loss function weights on three challenging medical imaging datasets in Section \ref{subsec4}.
\end{itemize}


\begin{table*}[!h]
\centering
\begin{tabular}{lllll}
\hline
\multirow{2}{*}{Encoder}    & \multirow{2}{*}{Technique} & \multicolumn{3}{c}{U-Net}                            \\
                            &                            & Baseline(w/o APP) & w/ Relu APP      & w/ Gelu APP      \\
                            \hline
\multirow{2}{*}{ResNet-18}  & DEDL                       & 0.4347           & 0.4608          & 0.4788          \\
                            & Ours                       & \textbf{0.5618}  & \textbf{0.5479} & \textbf{0.5582} \\
                            \hline
\multirow{2}{*}{ResNet-34}  & DEDL                       & 0.4774           & 0.4467          & 0.4983          \\
                            & Ours                       & \textbf{0.5306}  & \textbf{0.5571} & \textbf{0.5606} \\
                            \hline
\multirow{2}{*}{ResNet-50}  & DEDL                       & 0.3798           & 0.4187          & 0.3827          \\
                            & Ours                       & \textbf{0.5556}  & \textbf{0.5495} & \textbf{0.5597} \\
                            \hline
\multirow{2}{*}{ResNet-101} & DEDL                       & 0.3718           & 0.4074          & 0.4402          \\
                            & Ours                       & \textbf{0.5502}  & \textbf{0.5678} & \textbf{0.5497} \\
                            \hline
\end{tabular}
\caption{Performance comparison of DEDL \cite{singh2022data} and our approach on the dermatomyositis dataset for U-Net. We repeat all experiments five times with different seed values and report the IoU on the test in the 95\% CI (confidence interval). }
\label{table:ours_vs_DEDL_unet}
\end{table*}

\begin{table*}[!h]
\centering
\begin{tabular}{lllll}
\hline
\multirow{2}{*}{Encoder}    & \multirow{2}{*}{Technique} & \multicolumn{3}{c}{U-Net++}                          \\
                            &                            & Baseline(w/o APP) & w/ Relu APP      & w/ Gelu APP      \\
                            \hline
\multirow{2}{*}{ResNet-18}  & DEDL                       & 0.5274           & 0.4177          & 0.4707          \\
                            & Ours                       & \textbf{0.5622}  & \textbf{0.5679} & \textbf{0.5683} \\
                            \hline
\multirow{2}{*}{ResNet-34}  & DEDL                       & 0.3745           & 0.4535          & 0.4678          \\
                            & Ours                       & \textbf{0.5536}  & \textbf{0.5685} & \textbf{0.5633} \\
                            \hline
\multirow{2}{*}{ResNet-50}  & DEDL                       & 0.4236           & 0.4685          & 0.4422          \\
                            & Ours                       & \textbf{0.5742}  & \textbf{0.5698} & \textbf{0.5514} \\
                            \hline
\multirow{2}{*}{ResNet-101} & DEDL                       & 0.4311           & 0.4265          & 0.4467          \\
                            & Ours                       & \textbf{0.57}    & \textbf{0.5727} & \textbf{0.5692} \\
                            \hline
\end{tabular}
\caption{Similar to Table \ref{table:ours_vs_DEDL_unet}, in this table, we compare the performance comparison of DEDL \cite{singh2022data} and our approach on U-Net++. We report IoU scores averaged over five seed values in the 95\% confidence interval (CI) over the test set of the Dermatomyositis dataset in this table. }
\label{table:ours_vs_DEDL_unetpp}
\end{table*}
\section{Background}

\begin{table*}[!h]
    \centering
    \begin{tabular}{ccccc}
        \hline
         \multirow{2}{*}{\textbf{Dataset}} & \multirow{2}{*}{\textbf{ResNet}} & \multicolumn{3}{c}{ \textbf{U-Net} } \\
         & & Baseline(w/o APP) & w/ Relu APP & w/ Gelu APP \\
        \hline
        \multirow{4}{*}{\textbf{Dermofit}} 
        & ResNet18 & 0.7388 & \textbf{0.7477} & 0.7467  \\
        & ResNet34 & 0.7576 & \textbf{0.7633} & 0.7525  \\
        & ResNet50 & 0.7364 & 0.7338 & \textbf{0.7401}  \\
        & ResNet101 & 0.7252 & 0.7213 & \textbf{0.7258}  \\
        \hline
        \multirow{4}{*}{\textbf{Dermatomyositis}} 
        & ResNet18 & \textbf{0.5618} & 0.5479 & 0.5582  \\
        & ResNet34 & 0.5306 & 0.5571 & \textbf{0.5606}  \\
        & ResNet50 & 0.5556 & 0.5495 & \textbf{0.5597}  \\
        & ResNet101 & 0.5502 & \textbf{0.5678} & 0.5497  \\
        \hline
        \multirow{4}{*}{\textbf{ISIC2017}} 
        & ResNet18 & \textbf{0.6458} & 0.6252 & 0.6357  \\
        & ResNet34 & \textbf{0.6518} & 0.6227 & 0.6306  \\
        & ResNet50 & 0.605 & 0.5984 & \textbf{0.6207}  \\
        & ResNet101 & 0.6267 & \textbf{0.6325} & 0.5884  \\
        \hline
    \end{tabular} 
    \caption{In this table we present the IoU (Intersection over Union) on the test set averaged over five seed values (in 95\% CI) for U-Net trained with our proposed technique as mentioned in Section \ref{methods}.}
    \label{table:AEperformance}
    \end{table*}

\begin{table*}[!h]
    \centering
    \begin{tabular}{ccccc}
        \hline
         \multirow{2}{*}{\textbf{Dataset}} & \multirow{2}{*}{\textbf{ResNet}} &  \multicolumn{3}{c}{ \textbf{UNet++} } \\
         & &Baseline(w/o APP) & w/ Relu APP & w/ Gelu APP \\
        \hline
        \multirow{4}{*}{\textbf{Dermofit}} 
        & ResNet18 & \textbf{0.744} & 0.7408 & 0.7366 \\
        & ResNet34  & 0.754 & 0.7553 & \textbf{0.7599} \\
        & ResNet50  & 0.737 & \textbf{0.7408} & 0.7379 \\
        & ResNet101  & 0.7232 & \textbf{0.7264} & 0.7229 \\
        \hline
        \multirow{4}{*}{\textbf{Dermatomyositis}} 
        & ResNet18  & 0.5622 & 0.5679 & \textbf{0.5683} \\
        & ResNet34  & 0.5536 & \textbf{0.5685} & 0.5633 \\
        & ResNet50  & \textbf{0.5742} & 0.5698 & 0.5514 \\
        & ResNet101  & 0.57 & \textbf{0.5727} & 0.5692 \\
        \hline
        \multirow{4}{*}{\textbf{ISIC2017}} 
        & ResNet18  & 0.6096 & \textbf{0.6232} & 0.6005 \\
        & ResNet34  & \textbf{0.6583} & 0.6423 & 0.6548 \\
        & ResNet50 & 0.6103 & \textbf{0.6355} & 0.619 \\
        & ResNet101  & 0.6018 & \textbf{0.6164} & 0.6041 \\
        \hline
    \end{tabular}
    \caption{Similar, to table \ref{table:AEperformance} in this table we present IoU averaged over five seed values (in 95\% CI) for U-Net++.}
    \label{table:AEperformanceUnetpp}
    \end{table*}

    \begin{table*}[!htp]
    \centering
    
    \begin{tabular}{cccccc}
        \hline
         {\textbf{Supervision Level}} & {\textbf{Dataset}} &  { \textbf{Baseline(w/o APP)} } & \textbf{w/ Relu APP} & \textbf{w/ GeLu APP}\\
        \hline
        \multirow{3}{*}{\textbf{U-Net}} 
        & Dermofit & 0.7395 & \textbf{0.7415} &0.7413\\
        & Dermatomoyositis & 0.5496 & \textbf{0.5556} &0.5551 \\
        & ISIC2017 & \textbf{0.6323} & 0.6197 &0.6189\\
        \hline
        \multirow{3}{*}{\textbf{U-Net++}} 
        & Dermofit & 0.7396 & \textbf{0.7408} &0.7393\\
        & Dermatomoyositis & 0.565 & \textbf{0.5697} &0.5630\\
        & ISIC2017 & 0.6200 & \textbf{0.62935} &0.6196\\
        \hline
    \end{tabular} 
    \caption{Mean IoU in the 95\% confidence interval when averaged over the entire ResNet family for U-Net and U-Net++ on the dermofit, dermatomyositis, and the ISIC 2017 dataset. We observe that adding ReLU autoencoder as a post-processing unit improves performance for U-Net++ and, in almost all cases, for U-Net.}
    \label{app_overall}
    \end{table*}

Medical image segmentation separates the region of interest, usually a lesion, cells, or other anatomical region of interest, from the slide background. Traditional segmentation processes use pixel-level classification to group pixels into different categories; in the case of semantic segmentation, these would be background and foreground. But with the maturity of Convolutional Neural Networks(CNNs), Ronneberger \etal \cite{ronneberger2015u} introduced U-Net - an autoencoder-based architecture for biomedical segmentation. The U-Net consists of an encoder and a decoder architecture, where the encoder acts as a feature extractor, and the decoder learns the mask by using the extracted features as input. In addition, the decoder also incorporates the feature maps from the encoder to improve scaling up the representation to the image mask; these connections are called "Skip-connections." Following U-Net, a wealth of architectures have spawned:
U-Net++\cite{zhou2018unet++}, DeepLab \cite{chen2017deeplab}, DeepLabV3+ \cite{chen2018encoder}, and Feature Pyramid networks (FPN) \cite{lin2017feature}. All of these architectures build on the autoencoder architecture of U-Net with skip connections. To increase the receptive field of these architectures, various techniques have been introduced, such as dilated networks \cite{mehta2018espnet} and nesting architecture, as in the case of U-Net++\cite{zhou2018unet++}. Despite these advancements and complex architectures, U-Net remains the choice of architecture for medical image segmentation \cite{mehta2018ynet,singh2022data,van2022artificial}. 

\subsection{Application of Segmentation Techniques for Autoimmune diseases.}\label{ai-in-ai}

Stafford \etal \cite{stafford2020systematic} conducted a comprehensive survey to examine the application of AI in the context of autoimmune diseases. They observed the median size of autoimmune datasets is much smaller (99-540 samples per dataset) as compared to datasets pertaining to other medical modalities. The scarce available data poses a significant challenge in acquiring informative priors for artificial intelligence-based CAD approaches on these datasets resulting in sub-par performance. Furthermore, most methodologies for analyzing these datasets are antiquated and lack open-source availability\cite{singh2022data}. To overcome these shortcomings Van Buren \etal \cite{van2022artificial} proposed the use of U-Net for segmentation of whole slide images of dermatomyositis histopathology data and open-sourced their approach. Given the considerable size of the whole slide images (WSI) at 1408 $\times$ 1876, a tiling approach was employed to partition the WSI into smaller 256 $\times$ 256 images, with padding.

They also used a combination of Dice and Binary cross entropy loss to attenuate the problem of pixel distribution imbalance between the area surrounding the region of interest (background pixels) and the region of interest (foreground pixels). Due to this imbalance, the segmentation architecture tends to focus more on the area surrounding the region of interest than the region of interest for segmentation if unattended. Singh 
\& Cirrone \cite{singh2022data} further improved on this benchmark by using U-Net and introduced an ``Autoencoder Post Processing" (APP) technique.
The APP consists of stacked linear layers for the encoder and decoder. This makes the autoencoder much simpler than the convolution and skip connection based U-Net and U-Net++ architectures. After obtaining the mask from a U-Net or U-Net++, it is passed through the APP. Since, the autoencoder consists of only stacked linear layers it creates a noised version of the segmentation output from U-Net and U-Net++. A mean squared error is then calculated between the autoencoder's output and ground truth. During training, the model trained with the help of the MSE loss (calculated between the autoencoder output and the ground truth) and the cross entropy loss (calculated between the  U-Net/U-Net++ output and the ground truth). This helps the model learn a more diverse set of features.

The autoencoder is only used during the training process. Hence there is only a marginal increase in training time while the inference time remains constant. They studied ReLU and GELU as activation functions for the linear layers of APP and found that ReLU activations work better than GELU. For their choice of architecture, they used U-Net and U-Net++ (nested U-net) with Squeeze and Excitation \cite{hu2018squeeze} in the decoder for channel-level attention. To navigate the problem of pixel distribution imbalance between the area surrounding the region of interest (background pixels) and the region of interest (foreground pixels), they used pixel-distribution ratio weights in the cross-entropy loss. Wherein the background pixels used the ratio of background pixels to total pixels as weights, and similarly, the foreground pixels used the ratio of foreground pixels to total pixels as weights. With these changes, they were able to improve on state of the art on the dermatomyositis segmentation task \cite{van2022artificial} by around 5\%. They also suggested a change in evaluation metric from pixel accuracy to IoU (Intersection over Union) as pixel accuracy does not correctly represent the quality of the learned mask as opposed to the ground truth mask. Our study builds upon the foundation established by Singh \& Cirrone \cite{singh2022data} as a baseline.
Our proposed methodology demonstrates significant improvement in performance as compared to the state-of-the-art approach \cite{singh2022data}. We achieve an average improvement of 12.26\% for U-Net and 12.04\% for U-Net++, as elaborated in Section \ref{result_1}. Next, we benchmark our methodology on two complex skin lesion datasets in Section \ref{subsec3}. Furthermore, we investigate the impact of autoencoder for post-processing in Section \ref{subsec3} and the significance of loss function weights in Section \ref{subsec4}.

\section{Methodology}\label{methods}

We start with Singh \& Cirrone's \cite{singh2022data} approach on the dermatomyositis dataset. They use U-Net and U-Net++ as the choice of segmentation architecture with Squeeze and Excitation \cite{hu2018squeeze} in the decoder for channel-level attention. Similar to previous studies, our work focuses on semantic segmentation, where the goal is to categorize each pixel in an image into a class. For semantic segmentation, these classes would be - a region of interest (for example, cells in the case of the dermatomyositis dataset) and an area other than the region of interest or background (region other than the cells). For the encoder, we study the performance of the entire ResNet family of CNNs\cite{he2015deep}. We use an encoder depth of three, increasing the convolution filter size from 128, 256, to 512. We initialize the encoder with ImageNet pre-trained weights. In the decoder part of U-Net and U-Net++, we use a convolution channel scheme of 256, 128, and 64. For each decoder block, we also use batch normalization as well as squeeze-and-excitation channel excitation after the convolutional layer. As discussed in \cite{singh2022data,van2022artificial} and Section \ref{ai-in-ai}, the Dermatomyositis whole slide images contain a lot more pixels without cells (background) as opposed to with cells (foreground). To attenuate this imbalance in the distribution of pixels, we use \textbf{C}ross \textbf{D}istribution \textbf{W}eights (\textbf{CDW}) in the cross-entropy loss. Van Buren \etal \cite{van2022artificial} used random weights. In contrast, Singh \& Cirrone \cite{singh2022data} used a ratio of the pixel with cells to total pixels and a ratio of non-cell pixels to total pixels as weights for foreground and background, respectively, in the cross-entropy loss. We propose to swap the weights and instead use the ratio of the number of pixels not containing the cell to the total number of pixels as the weight for the foreground. Similarly, the weight of the background is the ratio of the number of pixels containing cells (foreground/object of interest) to the total number of pixels. This alternative weight assignment method aims to enhance the foreground representation. This intuition is very similar to focal loss \cite{lin2017focal}, wherein the misclassification of the minority class is penalized more than that of the minority class. To ensure our results are statistically significant, we conduct all experiments over five different seed values and report the mean values in the 95\% confidence interval (C.I) over the five runs. We present these results in Tables \ref{table:AEperformance} and \ref{table:AEperformanceUnetpp}. Based on our proposal, we observe an average improvement of 12.26\% for U-Net and 12.04\% for U-Net++. We further discuss these in Section \ref{result_1}. Additionally, we benchmark our approach and study the impact of autoencoder post-processing on two additional challenging dermatology-related datasets - ISIC 2017 and the dermofit dataset in Section \ref{subsec3}. Both datasets are challenging due to large intra-class variations and inter-class similarities as depicted in Figure \ref{fig:var_ex} and obscuration in \ref{fig:obs_ex}. Finally, we study the impact of using mean-frequency weights and compare the results with distribution-swapped weights for U-Net and U-Net++ over the ResNet family of encoders and three datasets in  Section \ref{subsec4}.


\begin{table*}[]
\centering
\begin{tabular}{ll|lll|lll}
\hline
\multirow{2}{*}{Encoder}    & \multirow{2}{*}{Technique} & \multicolumn{3}{c|}{U-Net}                                                                     & \multicolumn{3}{c}{U-Net++}                                                                   \\
                            &                            & \multicolumn{1}{l|}{Baseline*} & \multicolumn{1}{l|}{w/ ReLU APP}      & w/ GELU APP      & \multicolumn{1}{l|}{Baseline*} & \multicolumn{1}{l|}{w/ ReLU APP}      & w/ GELU APP      \\
                            \hline
\multirow{2}{*}{ResNet-18}  & CDW               & \multicolumn{1}{l|}{0.5618}           & \multicolumn{1}{l|}{0.5479}          & \textbf{0.5582} & \multicolumn{1}{l|}{0.5622}           & \multicolumn{1}{l|}{\textbf{0.5679}} & 0.5683 \\
                            & Mean Frequency             & \multicolumn{1}{l|}{\textbf{0.5645}}  & \multicolumn{1}{l|}{\textbf{0.5592}} & 0.5405          & \multicolumn{1}{l|}{\textbf{0.5852}}  & \multicolumn{1}{l|}{0.5603} & \textbf{0.5814} \\
                            \hline
\multirow{2}{*}{ResNet-34}  & CDW              & \multicolumn{1}{l|}{0.5306}           & \multicolumn{1}{l|}{\textbf{0.5571}} & 0.5606          & \multicolumn{1}{l|}{0.5536}           & \multicolumn{1}{l|}{0.5685} & 0.5633          \\
                            & Mean Frequency             & \multicolumn{1}{l|}{\textbf{0.5555}}  & \multicolumn{1}{l|}{0.5551}          & \textbf{0.5616} & \multicolumn{1}{l|}{\textbf{0.5729}}  & \multicolumn{1}{l|}{\textbf{0.5763}} & \textbf{0.57}   \\
                            \hline
\multirow{2}{*}{ResNet-50}  & CDW               & \multicolumn{1}{l|}{\textbf{0.5556}}  & \multicolumn{1}{l|}{0.5495}          & \textbf{0.5597} & \multicolumn{1}{l|}{\textbf{0.5742}}  & \multicolumn{1}{l|}{0.5698}          & 0.5514 \\
                            & Mean Frequency             & \multicolumn{1}{l|}{0.5512}           & \multicolumn{1}{l|}{\textbf{0.5652}} & 0.5585          & \multicolumn{1}{l|}{0.57}             & \multicolumn{1}{l|}{\textbf{0.5723}} & \textbf{0.5929} \\
                            \hline
\multirow{2}{*}{ResNet-101} & CDW              & \multicolumn{1}{l|}{0.5502}           & \multicolumn{1}{l|}{\textbf{0.5678}} & 0.5497          & \multicolumn{1}{l|}{0.57}             & \multicolumn{1}{l|}{\textbf{0.5727}} & 0.5692          \\
                            & Mean Frequency             & \multicolumn{1}{l|}{\textbf{0.5506}}  & \multicolumn{1}{l|}{0.5537}          & \textbf{0.5596} & \multicolumn{1}{l|}{\textbf{0.5892}}  & \multicolumn{1}{l|}{0.5678}          & \textbf{0.5773} \\
                            \hline
\end{tabular}
\caption{This table displays the average Intersection over Union (IoU) values obtained from five separate runs on the dermatomyositis test set, with their corresponding 95\% confidence intervals (CI). In this context, CDW refers to the utilization of cross-distribution weights in the calculation of cross-entropy loss. Here, Baseline* represents the use of the segmentation architecture with autoencoder for post-processing (APP).}
\label{mean_weight_cross_weight_Dermatomyositis}
\end{table*}

\begin{table*}[]
\centering
\begin{tabular}{ll|lll|lll}
\hline
\multirow{2}{*}{Encoder}    & \multirow{2}{*}{Technique} & \multicolumn{3}{c|}{U-Net}                                                                    & \multicolumn{3}{c}{U-Net++}                                                                  \\
                            &                            & \multicolumn{1}{l|}{Baseline*}       & \multicolumn{1}{l|}{w/ ReLU APP}      & w/ GELU APP      & \multicolumn{1}{l|}{Baseline*}       & \multicolumn{1}{l|}{w/ ReLU APP}      & w/ GELU APP      \\
                            \hline
\multirow{2}{*}{ResNet-18}  & CDW               & \multicolumn{1}{l|}{0.7388}          & \multicolumn{1}{l|}{0.7477}          & \textbf{0.7467} & \multicolumn{1}{l|}{0.744}           & \multicolumn{1}{l|}{0.7408}          & 0.7366          \\
                            & Mean Frequency             & \multicolumn{1}{l|}{\textbf{0.75}}   & \multicolumn{1}{l|}{\textbf{0.7498}} & 0.7377          & \multicolumn{1}{l|}{\textbf{0.7469}} & \multicolumn{1}{l|}{\textbf{0.7413}} & \textbf{0.7449} \\
                            \hline
\multirow{2}{*}{ResNet-34}  & CDW               & \multicolumn{1}{l|}{\textbf{0.7576}} & \multicolumn{1}{l|}{\textbf{0.7633}} & 0.7525          & \multicolumn{1}{l|}{0.754}           & \multicolumn{1}{l|}{0.7553} & \textbf{0.7599} \\
                            & Mean Frequency             & \multicolumn{1}{l|}{0.7533}          & \multicolumn{1}{l|}{\textbf{0.7633}} & \textbf{0.7535} & \multicolumn{1}{l|}{\textbf{0.7602}} & \multicolumn{1}{l|}{\textbf{0.7635}} & 0.7547          \\
                            \hline
\multirow{2}{*}{ResNet-50}  & CDW              & \multicolumn{1}{l|}{0.7364}          & \multicolumn{1}{l|}{\textbf{0.7338}} & \textbf{0.7401} & \multicolumn{1}{l|}{\textbf{0.737}}  & \multicolumn{1}{l|}{\textbf{0.7408}} & 0.7379          \\
                            & Mean Frequency             & \multicolumn{1}{l|}{\textbf{0.7379}} & \multicolumn{1}{l|}{0.731}           & 0.7385          & \multicolumn{1}{l|}{0.7362}          & \multicolumn{1}{l|}{0.7358}          & \textbf{0.7411} \\
                            \hline
\multirow{2}{*}{ResNet-101} & CDW              & \multicolumn{1}{l|}{\textbf{0.7252}} & \multicolumn{1}{l|}{0.7213}          & 0.7258          & \multicolumn{1}{l|}{\textbf{0.7232}} & \multicolumn{1}{l|}{\textbf{0.7264}} & 0.7229          \\
                            & Mean Frequency             & \multicolumn{1}{l|}{0.7212}          & \multicolumn{1}{l|}{\textbf{0.7242}} & \textbf{0.7236} & \multicolumn{1}{l|}{0.7156}          & \multicolumn{1}{l|}{0.7247}          & \textbf{0.7234} \\
                            \hline
\end{tabular}
\caption{Similar to Table \ref{mean_weight_cross_weight_Dermatomyositis}, in this table, we present the IoU averaged over five runs in 95\% confidence interval on the dermofit test set for U-Net and U-Net++. Like Table \ref{mean_weight_cross_weight_Dermatomyositis}, CDW represents the scenario when cross-distribution weights are used for the cross-entropy loss and Baseline* represents the use of the segmentation architecture with autoencoder for post-processing (APP).}
\label{mean_weight_cross_weight_Dermofit}
\end{table*}

\begin{table*}[]
\begin{tabular}{ll|lll|lll}
\hline
\multirow{2}{*}{Encoder}    & \multirow{2}{*}{Technique} & \multicolumn{3}{c|}{U-Net}                                                                     & \multicolumn{3}{c}{U-Net++}                                                                   \\
                            &                            & \multicolumn{1}{l|}{Baseline*} & \multicolumn{1}{l|}{w/ ReLU APP}      & w/ GELU APP      & \multicolumn{1}{l|}{Baseline*} & \multicolumn{1}{l|}{w/ ReLU APP}      & w/ GELU APP      \\
                            \hline
\multirow{2}{*}{ResNet-18}  & CDW              & \multicolumn{1}{l|}{\textbf{0.6458}}  & \multicolumn{1}{l|}{0.6252}          & \textbf{0.6357} & \multicolumn{1}{l|}{0.6096}           & \multicolumn{1}{l|}{\textbf{0.6232}} & 0.6005          \\
                            & Mean Frequency             & \multicolumn{1}{l|}{0.6257}           & \multicolumn{1}{l|}{\textbf{0.6394}} & 0.6307          & \multicolumn{1}{l|}{\textbf{0.6177}}  & \multicolumn{1}{l|}{0.6198}          & \textbf{0.6074} \\
                            \hline
\multirow{2}{*}{ResNet-34}  & CDW               & \multicolumn{1}{l|}{\textbf{0.6518}}  & \multicolumn{1}{l|}{0.6227}          & \textbf{0.6306} & \multicolumn{1}{l|}{\textbf{0.6583}}  & \multicolumn{1}{l|}{0.6423} & \textbf{0.6548} \\
                            & Mean Frequency             & \multicolumn{1}{l|}{0.6314}           & \multicolumn{1}{l|}{\textbf{0.6409}} & 0.6322          & \multicolumn{1}{l|}{0.6412}           & \multicolumn{1}{l|}{\textbf{0.6454}} & 0.6513          \\
                            \hline
\multirow{2}{*}{ResNet-50}  & CDW               & \multicolumn{1}{l|}{0.605}            & \multicolumn{1}{l|}{0.5984}          & 0.6207          & \multicolumn{1}{l|}{0.6103}           & \multicolumn{1}{l|}{\textbf{0.6355}} & 0.619           \\
                            & Mean Frequency             & \multicolumn{1}{l|}{\textbf{0.6396}}  & \multicolumn{1}{l|}{\textbf{0.6223}} & \textbf{0.6337} & \multicolumn{1}{l|}{\textbf{0.6354}}  & \multicolumn{1}{l|}{0.628}           & \textbf{0.646}  \\
                            \hline
\multirow{2}{*}{ResNet-101} & CDW             & \multicolumn{1}{l|}{\textbf{0.6267}}  & \multicolumn{1}{l|}{\textbf{0.6325}} & 0.5884          & \multicolumn{1}{l|}{0.6018}           & \multicolumn{1}{l|}{\textbf{0.6164}} & \textbf{0.6041} \\
                            & Mean Frequency             & \multicolumn{1}{l|}{0.6137}           & \multicolumn{1}{l|}{0.6283}          & \textbf{0.6175} & \multicolumn{1}{l|}{\textbf{0.6049}}  & \multicolumn{1}{l|}{0.6112}          & 0.6016      \\
                            \hline
\end{tabular}
\caption{In this table, we showcase the IoU results for U-Net and U-Net++ on the ISIC 2017 test set. The values represent the average IoU over five runs in the 95\% confidence interval. CDW, in this context, refers to the utilization of cross-distribution weights for cross-entropy loss, as demonstrated in Tables \ref{mean_weight_cross_weight_Dermatomyositis} and \ref{mean_weight_cross_weight_Dermofit}. Baseline* represents the use of the segmentation architecture with autoencoder for post-processing.}
\label{mean_weight_cross_weight_ISIC}
\end{table*}

    

\begin{table}[!htp]
    \centering
    \begin{tabular}{cccc}
        \hline
        {\textbf{Dataset}} &  { \textbf{CDW} } & \textbf{Median Frequency} \\
        \hline 
        Dermofit & 0.7395 & \textbf{0.7406} \\
        DM* & 0.5496 & \textbf{0.5564} \\
        ISIC2017 & 0.6197 & \textbf{0.6276} \\
        \hline
    \end{tabular} 
    \caption{In this table, we present the average IoU (in the 95\% confidence interval) over the ResNet family of encoders and the three paradigms (Baseline, with ReLU and GELU APP) for U-Net over the dermatomyositis, the dermofit, and ISIC 2017 test sets from Tables \ref{mean_weight_cross_weight_Dermatomyositis}, \ref{mean_weight_cross_weight_Dermofit} and \ref{mean_weight_cross_weight_ISIC} respectively. Here, DM* represents the Dermatomyositis dataset, and CDW represents Cross Distribution Weight.}
    \label{WeightVSMedFreqWeight_unet}
    \end{table}

\begin{table}[!htp]
    \centering
    
    \begin{tabular}{ccccc}
        \hline
        {\textbf{Dataset}} &  { \textbf{CDW} } & \textbf{Median Frequency} \\
        \hline
        
        Dermofit & 0.7396 & \textbf{0.7397} \\
        DM* & 0.565 & \textbf{0.5793} \\
        ISIC2017 & 0.62 & \textbf{0.6248} \\
        \hline
    \end{tabular} 
    \caption{Similar to Table \ref{WeightVSMedFreqWeight_unet}, In this table, we provide the average Intersection over Union (IoU) values (in 95\% CI) for the ResNet family for U-Net++ architecture on three different test sets: Dermatomyositis, Dermofit, and ISIC 2017.}
    \label{WeightVSMedFreqWeight_unetpp}
    \end{table}

\begin{table}[!h]
    \centering
    
    \begin{tabular}{ccc}
        \hline
        {\textbf{Dataset}} & \textbf{CDW} & \textbf{Median Frequency} \\
        \hline
        
        Dermofit & [0.3037, 0.6963] & [0.7180, 1.6466] \\
        
        DM* & [0.1479, 0.8521] & [0.5986, 3.0348] \\
        
        ISIC2017 & [0.2020, 0.7980] & [0.6265, 2.4755] \\
        \hline
    \end{tabular} 
    \label{table:weight_table}
    \caption{Cross-entropy weights used for experimentation as described in Section \ref{subsec3}. Here, DM* represents the Dermatomyositis dataset as described in Section \ref{dm_dataset} and CDW represents Cross Distribution Weight.}
\end{table}
    
\section{Experimental Details} \label{exp}

\subsection{Datasets}



We use a 70-10-20 split for the dermatomyositis and dermofit datasets. For the ISIC-2017 dataset, we use the same splits as used in the 2017 ISIC competition.

\paragraph{Dermatomyositis:}{\label{dm_dataset}} We use the same dataset as used in previous works on dermatomyositis segmentation \cite{singh2022data,van2022artificial}. To give an idea about the modality of the dataset, we show a random sample from the test set in Figure \ref{fig:isic_data}. The Dermatomyositis dataset is collected from 198 muscle biopsies collected from seven dermatomyositis patients.
These files are then stored in TIFF format. Each TIFF image contains eight slides that indicate the presence or absence of phenotypic markers by setting binary thresholds for each channel (1-DAPI, 2-CXCR3, 3-CD19, 4-CXCR5, 5-PD1, 6-CD4, 7-CD27, 8-Autofluorescence).
For segmentation, we used the DAPI-stained image. Each whole slide image was tiled into 480x480. We further expand on this in Section \ref{preprocessing}. This is a particularly challenging dataset due to the large number of fine-grained objects (cells) to be segmented per image, as discussed in Section \ref{intro}.

\paragraph{Dermofit \cite{Dermofit}:} As shown in Figure \ref{fig:isic_data},  the Dermofit dataset contains 1300 skin lesion RGB images. These data are taken with a high-quality SLR camera in controlled (ring flash) indoor illumination. 
The Dermofit dataset contains ten categories; each includes a different number of instances: Actinic Keratosis (AK): 45, Basal Cell Carcinoma (BCC): 239, Melanocytic Nevus / Mole (ML): 331, Squamous Cell Carcinoma (SCC) sample 88, Seborrhoeic Keratosis (SK): 257, Intraepithelial carcinoma (IEC): 78, Pyogenic Granuloma (PYO): 24, Haemangioma (VASC): 96, Dermatofibroma (DF): 65, Melanoma (MEL): 76. No two images in this dataset are of the same size, as a preprocessing step we interpolate all images to 480x480 and then resize to 224x224 to ensure uniformity.

\paragraph{ISIC Challenge 2017 Dataset, Lesion Segmentation Task \cite{codella2018skin}:} The International Skin Imaging Collaboration (ISIC) is a large publicly accessible dataset. We show a sample from the test set in Figure \ref{fig:isic_data}. In our case, we use the segmentation dataset from 2017 and use the original splits wherein 2,000 images were used as training, 150 images as validation, and 600 images as the test set. The ISIC 2017 and the Dermofit datasets described above are skin lesion datasets with high intra-class variability and inter-class similarities with obscuration areas of interest, as discussed in Section \ref{intro}.

\subsubsection{Common Implementation Details}We implemented all models in Pytorch \cite{NEURIPS2019_9015} using a single NVIDIA RTX-8000 GPU with 64 GB RAM and 3 CPU cores. All models are trained with an Adam optimizer with an initial learning rate (lr) of 3.6e-4 and a weight decay 1e-5.  
We use a cosine annealing scheduler with a maximum of 50 iterations and a minimum learning rate of 3.4e-4 to adjust the learning rate based on each epoch. We train all architectures for 50 epochs with batch size 16, followed by testing on a held-out set. We use IoU (Intersection over Union) as our evaluation metric on the test set. This aligns with previous work by Singh \& Cirrone \cite{singh2022data}. We repeat all experiments with five different seed values and report the mean value in the 95\% confidence interval in all tables. 
\subsubsection{Data-Preprocessing}\label{preprocessing}Images of the dermatomyositis dataset have a uniform size of 1408 $\times$ 1876; we tiled each image into 12 sub-images of size 480 $\times$ 480 inline with previous work \cite{singh2022data}.
In contrast, the Dermofit and the ISIC2017 datasets contain images of different sizes, i.e., no two images in the dataset are the same size. Additionally, since the other two datasets (dermofit and ISIC-2017) contain skin lesions, they have significantly denser and larger mask labels than the dermatomyositis dataset. Thus, a different image preprocessing step is applied to the latter two datasets: bilinear interpolation to 480 $\times$ 480 followed by a resize to 224 $\times$224. For augmentation, we use the same set of augmentation as used in Singh \& Cirrone's work \cite{singh2022data}, along with Red channel normalization or "Rnorm" \cite{rnorm} for all of our experiments.

\section{Results and Discussion}
\subsection{Improvement over the current state-of-the-art for Dermatomyositis WSI Segmentation \cite{singh2022data}}\label{result_1}
Following the methodology (Section \ref{methods}) and experimentation setup (Section \ref{exp}), we present the IoU averaged over five runs in the 95\% confidence interval on the test set for the Dermatomyositis dataset in Table \ref{table:ours_vs_DEDL_unet} for U-Net and in Table \ref{table:ours_vs_DEDL_unetpp} for U-Net++. We observe that our approach improves over Singh \& Cirrone's approach (DEDL) \cite{singh2022data} consistently over the entire ResNet family for baseline as well as with APP (both ReLU and GELU based) for both U-Net and U-Net++. When averaged over the ResNet family of encoders and the three paradigms (baseline approach without using autoencoders, ReLU autoencoders, and GELU-based autoencoders), we observed that our approach improves over the previous state-of-art \cite{singh2022data} for Dermatomyositis segmentation by 12.26\% and 12.04\% for U-Net and U-Net++ respectively.

\subsubsection{Impact of Incorporating Autoencoder Post-Processing.}\label{subsec3}

As described in Section \ref{ai-in-ai}, Singh and Cirrone \cite{singh2022data} introduced an "Autoencoder Post Processing" unit or APP after the main segmentation architecture. The purpose of this autoencoder was to provide a more noised version of the prediction from the U-Net or U-Net++. The mean square error loss between the noised output and the ground truth mask, along with the weighted-cross entropy loss between the output of the U-Net or U-Net++ and the ground truth, is optimized during training. This is depicted in Figure \ref{fig:DEDL}. They studied the impact of using APP with ReLU and GELU activations only on the Dermatomyositis dataset. In this section, with our improved approach as presented in Section \ref{methods}, we study the impact of adding APP on two additional challenging dermatology datasets. We present the IoU over the test set in the 95\% confidence interval over the ResNet family of encoders in Tables \ref{table:AEperformance} and \ref{table:AEperformanceUnetpp} for ISIC 2017 and the Dermofit dataset, respectively. Adding ReLu and GeLU-based APP improves performance over the baseline architecture (with no APP) for U-Net and U-Net++ in most cases for the Dermofit. To better understand the result, we average the IoU on the test set over the entire ResNet family for U-Net and U-Net++ and present the results in Table \ref{app_overall}. From Table \ref{app_overall}, we observe that the addition of APP, especially ReLU-based APP, does improve performance over the baseline (not using APP) in almost all cases for U-Net++ and U-Net. The addition of APP is did not improve performance only in the case of the ISIC-2017 dataset for U-Net.

\subsubsection{Impact of Cross-entropy loss weights}\label{subsec4}
In section \ref{methods}, we explained our rationale for switching from distribution-based weights to cross-distribution-based weights for the cross-entropy loss. In this section, we study the impact of changing the cross-entropy weights from cross-distribution-based weights to mean frequency weights\cite{DBLP:journals/corr/EigenF14}. The median frequency weight received by each class is derived from the reciprocal of the pixel ratio of a particular class, normalized by the median frequency of the class\cite{liu2019roadnet,liu2019deepcrack}. The median frequency and the cross-distribution weights, calculated over our three datasets, are mentioned in Table \ref{table:weight_table}. Mathematically, median frequency weights ($w_c$) are defined as follows: 
$w_c = \frac{\text{med\_freq}}{n_c}$. Here, $n_c$ is the number of pixels belonging to class $c$ in the training dataset, and $\text{med\_freq}$ is the median of the frequency of pixels belonging to each class in the dataset.\footnote{In our case, there are only two classes - foreground (region of interest) and background (area other than the region of interest).}
Where $n_c = \frac{\alpha}{\beta }$, here, $\alpha$  represents the number of pixels of a class, and $\beta$ represents the total number of pixels in images where the given class is present.
We compare the weights calculated by cross-distribution and median frequency in Table \ref{table:weight_table}. We provide the full comparative result of using cross-distribution weights and mean frequency over the three datasets in the 95\% confidence interval averaged over five seed values Tables \ref{mean_weight_cross_weight_Dermatomyositis}, \ref{mean_weight_cross_weight_Dermofit} and \ref{mean_weight_cross_weight_ISIC} for the Dermatomyositis, Dermofit, and the ISIC-2017 datasets, respectively. Additionally, to summarize these results, we present the average over the ResNet family and the three training paradigms (baseline without APP and APP with ReLU and GELU layers) in Tables \ref{WeightVSMedFreqWeight_unet} and \ref{WeightVSMedFreqWeight_unetpp} for U-Net and U-Net++, respectively. From these tables, we observe that median-frequency weights for cross-entropy loss improve performance over cross-distribution weights, although the improvement is marginal in almost all cases.

\section{Conclusion}

We observed that our approach of using Cross Distribution Weights (CDW) improved segmentation performance over the previous state-of-the-art approach for dermatomyositis segmentation \cite{singh2022data} by 12.26\% for U-Net and by 12.04\% for U-Net++ averaged over the ResNet family. Furthermore, adding APP (Autoencoder Post Processing) improves segmentation performance marginally in the case of dermatomyositis and dermofit datasets. In the case of the ISIC 2017 dataset, the addition of APP is only useful in the case of U-Net++. We have open-sourced our approach at \url{https://github.com/pranavsinghps1/Enhancing-Medical-Image-Segmentation}. We hope that our study and open-sourced approach will catalyze further research at the intersection of autoimmune diseases like dermatomyositis and the application of AI as well as for other dermatology-related datasets. This would help us better understand the immunology of autoimmune diseases and answer some of the critical research questions to develop improved healthcare solutions.\footnote{Potential negative societal impact: Autoimmune diseases are extremely heterogeneous; the dermatomyositis dataset used in our experiments is geographically restricted. Hence, this is a study of a particular variant. This study might or not be generalizable for other variants.  Hence application on a wider scale for real-life scenarios should only be trusted after clearance from the concerned health and safety governing bodies.}

\paragraph{Acknowledgements}We would like to thank NYU HPC
team for assisting us with our computational needs.

\newpage

{\small
\bibliographystyle{ieee_fullname}
\bibliography{egbib}

\begin{thebibliography}{10}\itemsep=-1pt

\bibitem{chen2017deeplab}
Liang-Chieh Chen, George Papandreou, Iasonas Kokkinos, Kevin Murphy, and Alan~L
  Yuille.
\newblock Deeplab: Semantic image segmentation with deep convolutional nets,
  atrous convolution, and fully connected crfs.
\newblock {\em IEEE transactions on pattern analysis and machine intelligence},
  40(4):834--848, 2017.

\bibitem{chen2018encoder}
Liang-Chieh Chen, Yukun Zhu, George Papandreou, Florian Schroff, and Hartwig
  Adam.
\newblock Encoder-decoder with atrous separable convolution for semantic image
  segmentation.
\newblock In {\em Proceedings of the European conference on computer vision
  (ECCV)}, pages 801--818, 2018.

\bibitem{codella2018skin}
Noel~CF Codella, David Gutman, M~Emre Celebi, Brian Helba, Michael~A Marchetti,
  Stephen~W Dusza, Aadi Kalloo, Konstantinos Liopyris, Nabin Mishra, Harald
  Kittler, et~al.
\newblock Skin lesion analysis toward melanoma detection: A challenge at the
  2017 international symposium on biomedical imaging (isbi), hosted by the
  international skin imaging collaboration (isic).
\newblock In {\em 2018 IEEE 15th international symposium on biomedical imaging
  (ISBI 2018)}, pages 168--172. IEEE, 2018.

\bibitem{cooper2009recent}
Glinda~S Cooper, Milele~LK Bynum, and Emily~C Somers.
\newblock Recent insights in the epidemiology of autoimmune diseases: improved
  prevalence estimates and understanding of clustering of diseases.
\newblock {\em Journal of autoimmunity}, 33(3-4):197--207, 2009.

\bibitem{ehrenfeld2020covid}
Michael Ehrenfeld, Angela Tincani, Laura Andreoli, Marco Cattalini, Assaf
  Greenbaum, Darja Kanduc, Jaume Alijotas-Reig, Vsevolod Zinserling, Natalia
  Semenova, Howard Amital, et~al.
\newblock Covid-19 and autoimmunity.
\newblock {\em Autoimmunity reviews}, 19(8):102597, 2020.

\bibitem{DBLP:journals/corr/EigenF14}
David Eigen and Rob Fergus.
\newblock Predicting depth, surface normals and semantic labels with a common
  multi-scale convolutional architecture.
\newblock {\em CoRR}, abs/1411.4734, 2014.

\bibitem{Dermofit}
Robert Fisher and Jonathan Rees.
\newblock Dermofit project datasets.
\newblock 2017.

\bibitem{galeotti2020autoimmune}
Caroline Galeotti and Jagadeesh Bayry.
\newblock Autoimmune and inflammatory diseases following covid-19.
\newblock {\em Nature Reviews Rheumatology}, 16(8):413--414, 2020.

\bibitem{he2015deep}
Kaiming He, Xiangyu Zhang, Shaoqing Ren, and Jian Sun.
\newblock Deep residual learning for image recognition. corr abs/1512.03385
  (2015), 2015.

\bibitem{hu2018squeeze}
Jie Hu, Li Shen, and Gang Sun.
\newblock Squeeze-and-excitation networks.
\newblock In {\em Proceedings of the IEEE conference on computer vision and
  pattern recognition}, pages 7132--7141, 2018.

\bibitem{lerner2015world}
Aaron Lerner, Patricia Jeremias, and Torsten Matthias.
\newblock The world incidence and prevalence of autoimmune diseases is
  increasing.
\newblock {\em Int J Celiac Dis}, 3(4):151--5, 2015.

\bibitem{lin2017feature}
Tsung-Yi Lin, Piotr Doll{\'a}r, Ross Girshick, Kaiming He, Bharath Hariharan,
  and Serge Belongie.
\newblock Feature pyramid networks for object detection.
\newblock In {\em Proceedings of the IEEE conference on computer vision and
  pattern recognition}, pages 2117--2125, 2017.

\bibitem{lin2017focal}
Tsung-Yi Lin, Priya Goyal, Ross Girshick, Kaiming He, and Piotr Doll{\'a}r.
\newblock Focal loss for dense object detection.
\newblock In {\em Proceedings of the IEEE international conference on computer
  vision}, pages 2980--2988, 2017.

\bibitem{liu2019roadnet}
Yahui Liu, Jian Yao, Xiaohu Lu, Menghan Xia, Xingbo Wang, and Yuan Liu.
\newblock Roadnet: Learning to comprehensively analyze road networks in complex
  urban scenes from high-resolution remotely sensed images.
\newblock {\em IEEE Transactions on Geoscience and Remote Sensing},
  57(4):2043--2056, 2019.

\bibitem{liu2019deepcrack}
Yahui Liu, Jian Yao, Xiaohu Lu, Renping Xie, and Li Li.
\newblock Deepcrack: A deep hierarchical feature learning architecture for
  crack segmentation.
\newblock {\em Neurocomputing}, 338:139--153, 2019.

\bibitem{matsoukas2022makes}
Christos Matsoukas, Johan~Fredin Haslum, Moein Sorkhei, Magnus S{\"o}derberg,
  and Kevin Smith.
\newblock What makes transfer learning work for medical images: feature reuse
  \& other factors.
\newblock In {\em Proceedings of the IEEE/CVF Conference on Computer Vision and
  Pattern Recognition}, pages 9225--9234, 2022.

\bibitem{mehta2018ynet}
Sachin Mehta, Ezgi Mercan, Jamen Bartlett, Donald Weaver, Joann Elmore, and
  Linda Shapiro.
\newblock Y-net: Joint segmentation and classification for diagnosis of breast
  biopsy images.
\newblock In {\em International Conference on Medical image computing and
  computer-assisted intervention}. Springer, 2018.

\bibitem{mehta2018espnet}
Sachin Mehta, Mohammad Rastegari, Anat Caspi, Linda Shapiro, and Hannaneh
  Hajishirzi.
\newblock Espnet: Efficient spatial pyramid of dilated convolutions for
  semantic segmentation.
\newblock In {\em Proceedings of the european conference on computer vision
  (ECCV)}, pages 552--568, 2018.

\bibitem{NEURIPS2019_9015}
Adam Paszke, Sam Gross, Francisco Massa, Adam Lerer, James Bradbury, Gregory
  Chanan, Trevor Killeen, Zeming Lin, Natalia Gimelshein, Luca Antiga, Alban
  Desmaison, Andreas Kopf, Edward Yang, Zachary DeVito, Martin Raison, Alykhan
  Tejani, Sasank Chilamkurthy, Benoit Steiner, Lu Fang, Junjie Bai, and Soumith
  Chintala.
\newblock Pytorch: An imperative style, high-performance deep learning library.
\newblock In {\em Advances in Neural Information Processing Systems 32}, pages
  8024--8035. Curran Associates, Inc., 2019.

\bibitem{ronneberger2015u}
Olaf Ronneberger, Philipp Fischer, and Thomas Brox.
\newblock U-net: Convolutional networks for biomedical image segmentation.
\newblock In {\em International Conference on Medical image computing and
  computer-assisted intervention}, pages 234--241. Springer, 2015.

\bibitem{sharma2023high}
Chetan Sharma and Jagadeesh Bayry.
\newblock High risk of autoimmune diseases after covid-19.
\newblock {\em Nature Reviews Rheumatology}, pages 1--2, 2023.

\bibitem{singh2022data}
Pranav Singh and Jacopo Cirrone.
\newblock A data-efficient deep learning framework for segmentation and
  classification of histopathology images.
\newblock In {\em European Conference on Computer Vision}, pages 385--405.
  Springer, 2022.

\bibitem{sriram2021covid}
Anuroop Sriram, Matthew Muckley, Koustuv Sinha, Farah Shamout, Joelle Pineau,
  Krzysztof~J Geras, Lea Azour, Yindalon Aphinyanaphongs, Nafissa Yakubova, and
  William Moore.
\newblock Covid-19 prognosis via self-supervised representation learning and
  multi-image prediction.
\newblock {\em arXiv preprint arXiv:2101.04909}, 2021.

\bibitem{stafford2020systematic}
IS Stafford, M Kellermann, E Mossotto, Robert~Mark Beattie, Ben~D MacArthur,
  and Sarah Ennis.
\newblock A systematic review of the applications of artificial intelligence
  and machine learning in autoimmune diseases.
\newblock {\em NPJ digital medicine}, 3(1):30, 2020.

\bibitem{rnorm}
Dang~N.H. Thanh, Uğur Erkan, V.B.~Surya Prasath, Vivek Kumar, and Nguyen~Ngoc
  Hien.
\newblock A skin lesion segmentation method for dermoscopic images based on
  adaptive thresholding with normalization of color models.
\newblock In {\em 2019 6th International Conference on Electrical and
  Electronics Engineering (ICEEE)}, pages 116--120, 2019.

\bibitem{van2022artificial}
Kayla Van~Buren, Yi Li, Fanghao Zhong, Yuan Ding, Amrutesh Puranik, Cynthia~A
  Loomis, Narges Razavian, and Timothy~B Niewold.
\newblock Artificial intelligence and deep learning to map immune cell types in
  inflamed human tissue.
\newblock {\em Journal of Immunological Methods}, 505:113233, 2022.

\bibitem{zhou2018unet++}
Zongwei Zhou, Md~Mahfuzur Rahman~Siddiquee, Nima Tajbakhsh, and Jianming Liang.
\newblock Unet++: A nested u-net architecture for medical image segmentation.
\newblock In {\em Deep learning in medical image analysis and multimodal
  learning for clinical decision support}, pages 3--11. Springer, 2018.

\end{thebibliography}
}

\end{document}